\documentclass[reprint,twocolumn,aps,prB,amsmath,amssymb,floatfix,superscriptaddress,longbibliography]{revtex4-1}
\usepackage{graphicx}
\usepackage{epsfig}
\usepackage{bm}
\usepackage{dcolumn}
\usepackage{color}
\usepackage{physics}
\usepackage{float}
\usepackage[colorlinks,urlcolor=blue,citecolor=blue,linkcolor=magenta]{hyperref}
\makeatletter
\newcommand*{\rom}[1]{\expandafter\@slowromancap\romannumeral #1@}
\makeatother
\usepackage{tikz}
\usepackage{subfigure}

\usepackage{diagbox}
\usepackage{booktabs}
\usepackage{multirow}
\usepackage{makecell}
\usepackage{amsfonts}
\usepackage{amssymb}
\usepackage{graphicx}
\usepackage{dcolumn}
\usepackage{bm}
\usepackage{amsmath}
\usepackage{graphicx}
\usepackage{subfigure}
\usepackage{float}
\usepackage{color}

\begin{document}

\preprint{APS/123-QED}
\preprint{This line only printed with preprint option}

\title{Emergent extended states in an unbounded quasiperiodic lattice}

\author{Jia-Ming Zhang}
\affiliation {Key Laboratory of Atomic and Subatomic Structure and Quantum Control (Ministry of Education), Guangdong Basic Research Center of Excellence for Structure and Fundamental Interactions of Matter, School of Physics, South China Normal University, Guangzhou 510006, China}
\affiliation {Guangdong Provincial Key Laboratory of Quantum Engineering and Quantum Materials, Guangdong-Hong Kong Joint Laboratory of Quantum Matter, Frontier Research Institute for Physics, South China Normal University, Guangzhou 510006, China}

\author{Shan-Zhong Li}
\affiliation {Key Laboratory of Atomic and Subatomic Structure and Quantum Control (Ministry of Education), Guangdong Basic Research Center of Excellence for Structure and Fundamental Interactions of Matter, School of Physics, South China Normal University, Guangzhou 510006, China}
\affiliation {Guangdong Provincial Key Laboratory of Quantum Engineering and Quantum Materials, Guangdong-Hong Kong Joint Laboratory of Quantum Matter, Frontier Research Institute for Physics, South China Normal University, Guangzhou 510006, China}

\author{Shi-Liang Zhu}
\affiliation {Key Laboratory of Atomic and Subatomic Structure and Quantum Control (Ministry of Education), Guangdong Basic Research Center of Excellence for Structure and Fundamental Interactions of Matter, School of Physics, South China Normal University, Guangzhou 510006, China}
\affiliation {Guangdong Provincial Key Laboratory of Quantum Engineering and Quantum Materials, Guangdong-Hong Kong Joint Laboratory of Quantum Matter, Frontier Research Institute for Physics, South China Normal University, Guangzhou 510006, China}
\affiliation{Quantum Science Center of Guangdong-Hong Kong-Macao Greater Bay Area, Shenzhen, China}

\author{Zhi Li}
\email[Corresponding author: ]{lizphys@m.scnu.edu.cn}
\affiliation {Key Laboratory of Atomic and Subatomic Structure and Quantum Control (Ministry of Education), Guangdong Basic Research Center of Excellence for Structure and Fundamental Interactions of Matter, School of Physics, South China Normal University, Guangzhou 510006, China}
\affiliation {Guangdong Provincial Key Laboratory of Quantum Engineering and Quantum Materials, Guangdong-Hong Kong Joint Laboratory of Quantum Matter, Frontier Research Institute for Physics, South China Normal University, Guangzhou 510006, China}

\date{\today}

\begin{abstract}
Previous studies have established that quasiperiodic lattice models with unbounded potentials can exhibit localized and multifractal states, yet preclude the existence of extended states. In this work, we introduce a quasiperiodic system that incorporates both unbounded potentials and unbounded hopping amplitudes, where extended states emerge as a direct consequence of the unbounded hopping terms overcoming the localization constraints imposed by the unbounded potential, thereby facilitating enhanced particle transport. By using Avila's global theory, we derive analytical expressions for the phase boundaries, with exact results aligning closely with numerical simulations. Intriguingly, we uncover a hidden self-duality in the proposed model by establishing a mapping to the Aubry-Andr\'{e} model, revealing a profound structural connection between these systems.
\end{abstract}

\maketitle

\section{Introduction}

Inspired by electron spin resonance experiments on silicon conducted by G. Feher and E.A. Gere, P.W. Anderson formulated the Anderson localization theory to elucidate metal-insulator phase transitions in disordered systems~\cite{PWAnderson1958}. Subsequent scaling theory analyses demonstrated that in one-dimensional (1D) and two-dimensional (2D) systems with random disorder potentials, even infinitesimal disorder induces complete wavefunction localization~\cite{EAbrahams1979, PALee1985, BHetenyi2021}. In contrast, three-dimensional (3D) systems with weak disorder exhibit a coexistence of extended and localized states. This implies the existence of a critical energy threshold, termed the mobility edge, that separates these distinct phases~\cite{FEvers2008, ALagendijk2009, NMott1987}. Advancements in semiconductor doping technology have spurred various applications of mobility edge phenomena. Notably, recent researches in thermoelectric devices, which have strong thermoelectric effects, exemplify practical implementations  in mobility edge engineering~\cite{RWhitney2014, YamamotoK2017, CChiaracane2020}.

Quasiperiodic systems, which interpolate between random disorder and strict periodicity, have garnered significant attention due to their simple mathematical form and analytical tractability. These systems serve as pivotal frameworks for investigating Anderson transitions and mobility edges in low-dimensional systems~\cite{DJThouless1988, MKohmoto1983, MKohmoto2008, XCai2013, GRoati2008, YLahini2009, DTanese2014, HPLuschen2018, FAAn2018,FAAn2021, YWang2022a, HLi2023,TYLi2024}. Among these, the Aubry-Andr\'{e} (AA) model is a paradigmatic 1D quasiperiodic system. Its self-dual symmetry enables exact analytical determination of the critical point governing the Anderson transition~\cite{SAubry1980, PGHarper1955, MGon?alves2022}. In this model, eigenstates adopt extended characteristics below the critical disorder strength and become fully localized above it, with multifractal behavior emerging precisely at the critical point. Recent advances have expanded the AA model's utility by introducing modifications such as tailored quasiperiodic potentials~\cite{SDSarma1988, SDSarma1990, ZLu2022, YWang2020a,SLZhu2013,SGaneshan2015, HYao2019, XLi2020, TLiu2022, XPLi2016, XLi2017, BFZhu2023, EWLiang2023}, engineered hopping terms~\cite{JBiddle2010, JBiddle2011, XDeng2019, XXia2022, MGon2023, XCZhou2023a}, and non-Hermitian extensions~\cite{YJZhao2025, SZLi2024a, SZLi2024b, GJLiu2024, SZLi2024c, SLJiang2023, DWZhang2020a, DWZhang2020b, HJiang2019, JLDong2025, QLin2022, TLi2022}. These extensions have revealed rich phenomena, including the emergence of mobility edges, thereby broadening the scope of quantum phase transitions in disordered systems.

Furthermore, quasiperiodic systems can host a third distinct state that is neither extended nor localized, called multifractal state. This state derives its name from the multifractal structure of its eigenstate wavefunctions. In terms of dynamical properties and energy spectrum statistics, multifractal states exhibit marked differences from their extended or localized counterparts, as extensively discussed in studies of quasiperiodic systems~\cite{MGon2023, HLi2023, YHatsugai1990, JHHan1994, YTakada2004, FLiu2015, JWang2016}. Recent years have seen growing interest in multifractal states and multifractal-enriched mobility edges (MMEs)~\cite{SZLi2025}. Among systems capable of hosting MMEs, those with unbounded quasiperiodic potentials have garnered significant attention due to their rich spectral behavior~\cite{SJitomirskaya2017, FYang2019, JWang2022, YCZhang2022, XPJiang}.
Mathematically, unbounded quasiperiodic potentials are described by meromorphic functions, where singularities at specific sites constrain the spatial extension of wavefunctions. The Simon-Spencer theory establishes that such unbounded potentials disrupt the absolute continuity of the energy spectrum, leading to either a piecewise continuous spectrum with singular points or a purely discrete point spectrum~\cite{BSimon1989, SLonghi2023}. The former corresponds to multifractal states, while the latter aligns with localized states.

\renewcommand{\arraystretch}{2} 
\begin{table}[ht]
\caption{Possible emergent states in an unbounded system.}

\begin{tabular}{|c| c |c |}
\hline\hline
\multicolumn{1}{|c|}{\diagbox[width=3.2cm,height=1.3cm]{Hopping}{Potential}} & \makecell{Bound\\(HF$^*$)\\} &\makecell{Unbound\\(MF)\\}\\

\hline
\makecell{Normal (Constant)\\}&Extended~\cite{{SAubry1980}}&Multifractal~\cite{{SGaneshan2015}}
\\

\hline
\makecell{Unbound\\(MF)\\}&/&\makecell{Extended\\(Present work)}\\
\hline

\end{tabular}
\label{Tab1}

\vspace{0.5em}
$^*$HF: Holomorphic Function;   MF: Meromorphic Function
\end{table}

In previously studied unbounded systems, the potential energy typically exhibits an unbounded functional form, while the hopping term remains constant. This raises a critical question: Can extended states in unbounded systems be preserved by modifying the functional structure of the hopping term? To address this, we introduce a quasiperiodic lattice model featuring bothunbounded hopping and unbounded potential. Our key finding is that extended states can indeed emerge in such systems - the unbounded hopping counteracts the localization effects induced by the unbounded potential, effectively shielding delocalized states from suppression (see Tab.~\ref{Tab1}). Specifically, we demonstrate that unbounded hopping terms can dominate over the localization constraints imposed by the unbounded potential, enabling robust particle transport.
Using Avila's global theory, we derive an analytical expression for the mobility edges, and the results show excellent agreement with numerical simulations. Strikingly, we uncover a hidden self-duality in the model by mapping it to the Aubry-Andr\'{e} Hamiltonian, thereby revealing a deep structural correspondence between these systems. This duality elucidates the intricate competition between unbounded potentials and hopping amplitudes in quasiperiodic lattices, offering new insights into localization-delocalization transitions in non-periodic systems.

The rest of this manuscript is organized as follows:
In Section~\ref{Sec2}, we introduce the theoretical framework and define the key observables  to our analysis. Section~\ref{Sec3} presents the main results, including detailed numerical and analytical results. The underlying protection mechanism responsible for stabilizing the extended state is examined in Section~\ref{Sec4}. Finally, we summarize our conclusions in Section~\ref{Sec5}.

\section{Model and key observables}\label{Sec2}
\begin{figure}[htbp]
\centering
\includegraphics[width=8.5cm]{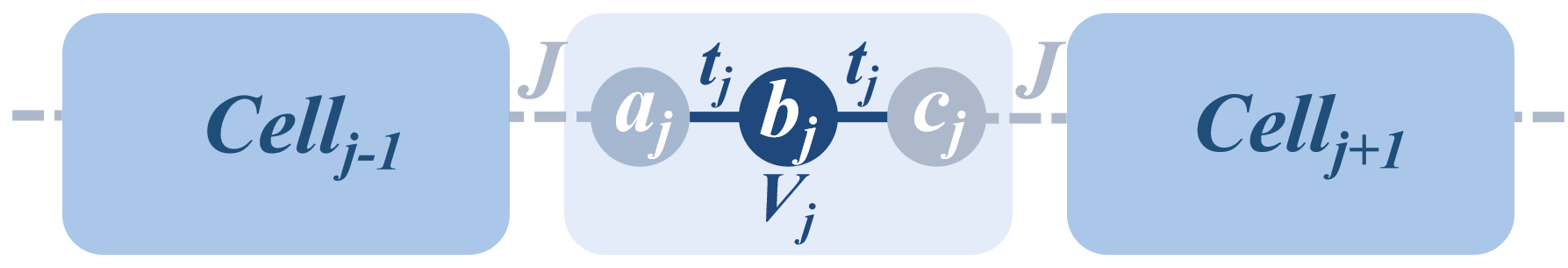}
\caption{Schematic diagram of the model in Eq.~\eqref{E1}.}
\label{F1}
\end{figure}

We propose an exactly solvable lattice model with both potential and hopping terms of unbounded quasiperiodic structure. The corresponding Hamiltonian reads,
\begin{equation}\label{E1}
H=\sum_{j=1}^{L}(Jc_{j-1}^\dag a_{j}+t_ja_{j}^\dag b_{j}+t_jb_{j}^\dag c_{j}+H.c.)+\sum_{j=1}^LV_jb_j^\dag b_j
\end{equation}
where
\begin{equation}\label{E2}
V_j=\frac{\lambda}{\cos^2(2\pi\alpha j+\theta)}~~\&~~t_j=\frac{t}{\cos(2\pi\alpha j+\theta)},
\end{equation}
with $a_{j}^{\dagger}$, $b_{j}^{\dagger}$ and $c_{j}^{\dagger}$ ($a_{j}$, $b_{j}$ and $c_{j}$) denoting creation (annihilation) operators on $a$, $b$ and $c$ sublattices of the $j$-th site, respectively. $\lambda$ is the strength of the quasiperiodic potential. $t_j$ and $J$ correspond to the strength of intra- and extra-hopping of unit cells, respectively~(see Fig.~\ref{F1}). $\alpha$ and $\theta$ are the strength of quasiperiodic parameter and global phase. Without loss of generality, we set $\theta=0$ and $\alpha=\frac{\sqrt{5}-1}{2}$ in the following calculations.

On the one hand, the fractal dimension~\cite{YWang2020a, YWang2022b}, as the core quantity reflecting the localization properties, is defined as

\begin{equation}
\Gamma_{\beta}=-\lim_{L\rightarrow\infty}\frac{\ln\xi_{\beta}}{\ln L},
\end{equation}
where $\xi_{\beta}=\sum_{j=1}^{L}|\psi_{j}(\beta)|^4$ is the inverse participation ratio and $\psi_{j}(\beta)$ is the amplitude of the $\beta$-th eigenstate at the $j$-th site. $\Gamma=0$ ($\Gamma=1$) corresponds to a localized (extended) state, while $0<\Gamma<1$ corresponds to a multifractal state.




On the other hand, level spacings is an powerful indicator to distinguish the extended, localized and multifractal states. One can define the even-odd (odd-
even) level spacings of the eigenvalues, i.e., $\delta^{e-o}_n=E_{2n}-E_{2n-1}$ ($\delta^{o-e}_n=E_{2n+1}-E_{2n}$), where $n=1,~2,~3,\cdots$~\cite{APadhan2022, YZhang2022, MSarkar2021, XDeng2019, RQi2023}. $E_{2n}$ ($E_{2n-1}$) denotes the even (odd) eigenenergy in ascending order of the eigenenergy spectrum. Since the extended single-particle states exhibit a doubly-degenerate spectrum, the corresponding level spacing $\delta^{e-o}_n\approx0$. A gap between $\delta^{e-o}_n$ and $\delta^{o-e}_n$ appears. As for the localized state, the corresponding $\delta^{e-o}_n$ and $\delta^{o-e}_n$ are almost same and the gap no longer exists. Since the properties are in between, the corresponding level spacing $\delta^{e-o}_n$ and $\delta^{o-e}_n$ of amultifractal states exhibits the scatter distributed behavior.

We have summarized the properties of two key observables in Tab.~\ref{Tab2}.

\begin{table}[tbhp]
\renewcommand{\arraystretch}{2}
	\centering
 	\caption{Key indicators of states' localization feature}
	\begin{tabular}{ c c c c}
		\hline\hline
		States    &             Fractal Dimension      &             Level Spacing (LS)       &    Gap of LS
 \\  \hline
		Ext.      & $\Gamma=1$            & $0\approx\delta^{e-o}<\delta^{o-e}$   & \checkmark  \\
		Loc.      & $\Gamma=0$            & $0<\delta^{e-o}\approx\delta^{o-e}$   & $\times$    \\
		Mul.      & $0<\Gamma<1$          &       Scatter Distribution            & /    \\
\hline\hline
	\end{tabular}
	\label{Tab2}
\end{table}

\section{Main Results}\label{Sec3}

\begin{figure*}[htbp]
\centering
\includegraphics[width=17cm]{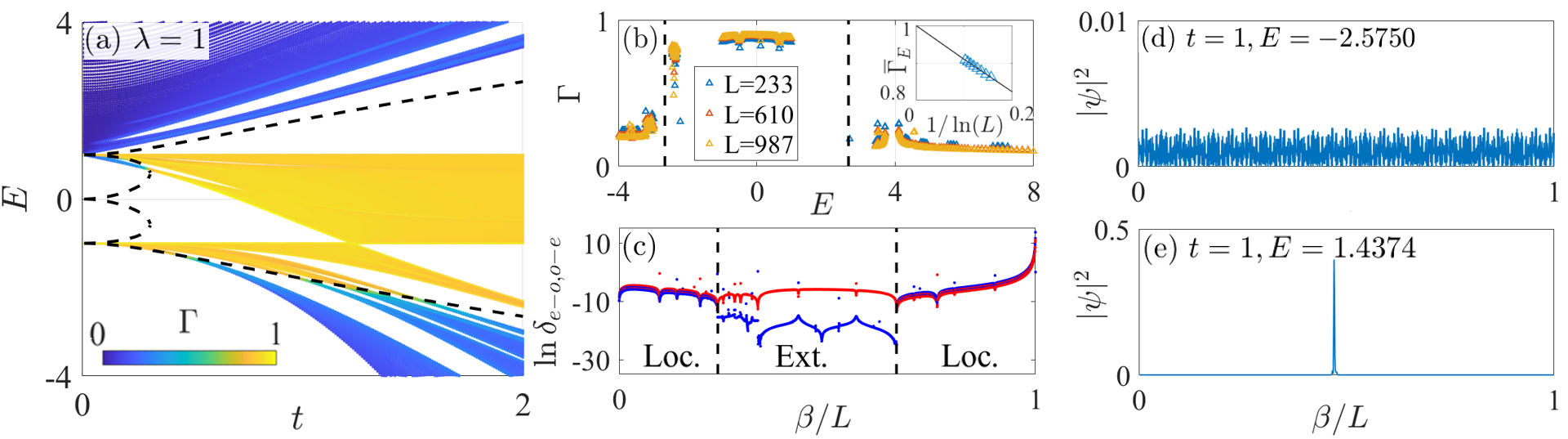}
\caption{Phase diagrams of fractal dimensions in $t$-$E$ plane with $\lambda=1$ (a). (b) The $t=2$ cross profile of (a) for different system size. The inset reveals the case of the thermodynamic limit by inverse interpolation. (c) The level spacings $\ln\delta_{e-o}$ (blue) and $\ln\delta_{o-e}$ (red) under the condition of $\lambda=1$ and $t=2$. Density distribution of eigenstate in extended (d) and localized (e) region of (a). Throughout, we set $J=1$ as the energy unit. $L=987$ for (a), (d) and (e), while $L=6765$ for (c).}
\label{F2}
\end{figure*}

Through numerical calculations, we determine the fractal dimension and energy level spacing associated with the Hamiltonian in Eq.~\eqref{E1} for varying parameters. The resulting phase diagram and level spacing distributions are illustrated in Fig.~\ref{F2}.

As demonstrated in Fig.~\ref{F2}(a), the system exhibits only localized and extended states, with no evidence of multifractal behavior. To validate the robustness of these findings across varying system sizes, we conducted a finite-size analysis. This analysis confirms the absence of multifractal states in the model described by Eq.~\eqref{E1} (Fig.~\ref{F2}(b)). Furthermore, the level spacing statistics reveal a clear dichotomy between localized and extended phases. Notably, the transition between even-odd and odd-even level spacing gaps is distinctly binary, either a pronounced gap exists or it is entirely absent (Fig.~\ref{F2}(c)). Crucially, the scattered distribution characteristic of multifractal states is not observed here, reinforcing the conclusion that such states are absent in this system. Additionally, the level spacing results confirm that the extended phase is robustly protected.

Finally, the spatial profiles of wave functions in distinct phases are depicted in Fig.~\ref{F2}(d)-(e), further illustrating the sharp contrast between localized and extended states.

\subsection{The exact solution of the mobility edge}\label{Sec2a}
Now, we turn to derive the exact expression of the mobility edge. From Eq.~\eqref{E1}, one can obtain the following eigenequation set,
\begin{equation}\label{E4}
\left\{\begin{matrix}
E\psi_{a,j}=J\psi_{c,j-1}+t_j\psi_{b,j},\\
E\psi_{b,j}=V_j\psi_{b,j}+t_j(\psi_{aj}+\psi_{c,j}),\\
E\psi_{c,j}=t_j\psi_{b,j}+J\psi_{a,j+1}.
\end{matrix}\right.
\end{equation}
The corresponding transition matrix reads
\begin{equation}
\begin{split}
T_{j}(\theta)&=
\left(
\begin{array}{cc}
\frac{E}{J}&-\frac{t_j}{J}\\
1&0
\end{array}
\right)
\left(
\begin{array}{cc}
\frac{E-V_j}{t_j}&-1\\
1&0
\end{array}
\right)
\left(
\begin{array}{cc}
\frac{E}{t_j}&-\frac{J}{t_j}\\
1&0
\end{array}
\right)\\
&=
\left(
\begin{array}{cc}
\frac{E^2(E-V_j)}{Jt_j^2}-\frac{2E}{J}&-\frac{E(E-V_j)}{t_j^2}+1\\
\frac{E(E-V_j)}{t_j^2}-1&-\frac{J(E-V_j)}{t_j^2}
\end{array}
\right).
\end{split}
\end{equation}
Substituting the specific forms~(Eq.~\eqref{E2}) of $V_j$ into the above expressions, one can get
\begin{equation}
T_{j}(\theta)=
\left(
\begin{array}{cc}
\frac{E^3}{Jt^2}C_j^2-\frac{E^2\lambda}{Jt^2}-\frac{2E}{J}&-\frac{E^2}{t^2}C_j^2+\frac{E\lambda}{t^2}+1\\
\frac{E^2}{t^2}C_j^2-\frac{E\lambda}{t^2}-1&-\frac{JE}{t^2}C_j^2+\frac{J\lambda}{t^2}
\end{array}
\right),
\end{equation}
where $C_j=\cos(2\pi\alpha j+\theta)$. In order to obtain an analytical expression for the Lyapunov exponent, we employ Avila's global theory of one-frequency analytical $SL(2,\mathbb{C})$ cocycle~\cite{{Avila2015}}. By letting $\theta\to\theta+i\epsilon$ and under the case of large $\epsilon$ limit, we obtain
\begin{equation}
T_{j}(\theta)=
\frac{e^{-i2\pi\alpha j}e^{\vert\epsilon\vert}}{4}
\left(
\begin{array}{cc}
\frac{E^3}{Jt^2}&-\frac{E^2}{t^2}\\
\frac{E^2}{t^2}&-\frac{JE}{t^2}
\end{array}
\right)+\mathcal{O}(1).
\end{equation}

Then, one can get the Lyapunov exponent,
\begin{equation}
\begin{split}
\gamma(E)&=\lim\limits_{L\to\infty}\frac{1}{L}\ln\Arrowvert\prod_{j=1}^{L}T_{j}(\theta)\Arrowvert\\
&=\ln\vert\frac{\frac{E^3}{Jt^2}-\frac{JE}{t^2}}{4}\vert+\epsilon+\mathcal{O}(1).
\end{split}
\end{equation}
According to global theory, $\gamma(E)$ is a convex piecewise linear function with respect to $\epsilon$. In large
$\epsilon$ limit, the slope of $\gamma$ is always one, which means that $\mathcal{O}(1)$ term can be ignored. Since Lyapunov exponent's expression is an even function. Then, the energy $E$ belongs to the spectrum of Hamiltonian~\eqref{E1}, we have
\begin{equation}
\begin{split}
\gamma(E)=\max\{\ln\vert\frac{\frac{E^3}{Jt^2}-\frac{JE}{t^2}}{4}\vert+|\epsilon|,~0\}.
\end{split}
\end{equation}
Since $\epsilon$ affects Lyapunov exponents only quantitatively (increasing the overall value), not qualitatively, it is safe to drop the $\epsilon$ term from the final expression. Finally, we obtain
\begin{equation}
\begin{split}
\gamma(E)=\max\{\ln\vert\frac{\frac{E^3}{Jt^2}-\frac{JE}{t^2}}{4}\vert,~0\}.
\end{split}
\end{equation}
From the above expression, one can find that when the input value in natural logarithm function is greater than one, the Lyapunov exponent is larger than zero, which means $\ln|1|$ is the critical point. Based on this, one can obtain the analytical expression of the mobility edge,
\begin{equation}\label{E11}
\vert E^3-J^2E\pm4Jt^2\vert=0.
\end{equation}
The analytical expression agree with numerical results well~(see black dashed line in Fig.~\ref{F2}(a)).

\subsection{The hidden self-duality}\label{Sec2b}
Further analysis reveals a hidden self-duality of the model~\eqref{E1}. Since this helps to deeper understand from another perspective why unbounded hopping can protect the extended state, we will clarify this in detail in this section.

First, let us consider each unit cell as an effective lattice point. Then, Eq.~\eqref{E4} can be mapped to a equivalent eigen equation, i.e.,
\begin{equation}
\begin{split}
E\psi_{b,j}=&V_j\psi_{b,j}+\frac{2t_j^2}{E-\frac{J^2}{E}}\psi_{b,j}\\
&+\frac{Jt_j}{E^2-J^2}(t_{j+1}\psi_{b,j+1}+t_{j-1}\psi_{b,j-1}).
\end{split}
\end{equation}
Let
\begin{equation}
\tilde{\psi}_{b,j}=\frac{1}{E^2-J^2}t_j\psi_{b,j}.
\end{equation}
One can obtain
\begin{equation}\label{E14}
2E\tilde{\psi}_{b,j}=\frac{E(E^2-J^2)}{t_j^2}\tilde{\psi}_{b,j}-\frac{V_j}{t_j^2}\tilde{\psi}_{b,j}+J(\tilde{\psi}_{b,j+1}+\tilde{\psi}_{b,j-1}).
\end{equation}
Note that, $t_j$ and $V_j$ appear in the potential function after mapping, while $J$ is the hopping strength of the effective model. Then, one can obtain a effective Hamiltonian~\cite{MGong2024}
\begin{equation}\label{E15}
H_{\text{eff}}=\sum_{j=1}^{L-1}J(c_{j}^{\dagger}c_{j+1}+H.c.)+\sum_{j=1}^{L}\mathcal{V}_{\text{eff},j}c_{j}^{\dagger}c_{j}
\end{equation}
where
\begin{equation}
\mathcal{V}_{\text{eff},j}=\frac{1}{t_j^2}[E(E^2-J^2)-V_j].
\end{equation}
Note that, the expression~\eqref{E15} has the structure of the AA model. Since $t_j$ satisfy Eq.~\eqref{E2}, by using the self-duality of the AA model, we obtain $\frac{E(E^2-J^2)}{t^2}=\pm 4J$. Based on this, one can get the same analytical expression of the mobility edge as Eq.~\eqref{E11}. From the expression, one can also find that model~\eqref{E1} can be mapped into a AA model. The intra-hopping of unit cells can be mapped to the on-site potential term in the effective model, while only unit cells' extra-hopping $J$ serves as the effective hopping in the mapped model. Although the original structure of model~\eqref{E1} appears to have no self-duality, it actually has a hidden self-duality.

\section{The mechanism behind the extended state being protected}\label{Sec4}

In this section, we present a detailed analysis demonstrating that extended states can persist in systems with unbounded potentials. While previous studies suggest that unbounded potentials suppress extended states and induce multifractal behavior~\cite{SJitomirskaya2017, FYang2019, JWang2022, YCZhang2022, XPJiang}, the precise mechanism driving multifractal formation remains unclear. By analyzing existing models, we identify a common feature in systems hosting multifractal states: the presence of specific sites that impede wavefunction propagation. For instance, singularities in unbounded potentials, acting as infinite barriers, restrict wavefunction delocalization, thereby enabling multifractality. Notably, systems with bounded quasiperiodic hopping (containing zeros) can be mapped to unbounded potential models, explaining why multifractal states also arise in such cases.

The model studied here incorporates both unbounded hopping and unbounded potential terms. Crucially, these terms are meromorphic functions sharing singularities at identical sites. Upon mapping, the unbounded potential (numerator) cancels the unbounded hopping (denominator), reducing the system to a conventional AA model. This restoration of the AA framework revives extended states. Furthermore, when the hopping term exhibits a squared relationship with the quasiperiodic potential structure, the mapping invariably recovers the AA model, ensuring extended-state protection in this model class. We thus conjecture that any system where the hopping structure cancels the unbounded potential post-mapping will preserve extended states.

Analytically, we uncover a hidden self-duality in the model. Mapping transforms the system into an AA model with a single mobility edge separating extended and localized phases. From Eq.~\eqref{E15}, the intra-unit-cell hopping introduces an effective quasiperiodic potential $\mathcal{V}_j\propto\frac{1}{t_j^2}$. Here, poles (zeros) in the unbounded (bounded) hopping map to zeros (poles) in the potential. This duality explains why multifractal states emerge in quasiperiodic hopping models with zeros\cite{XCZhou2023a}: such systems are dual to unbounded potential models.

\begin{figure}[htbp]
\centering
\includegraphics[width=8.5cm]{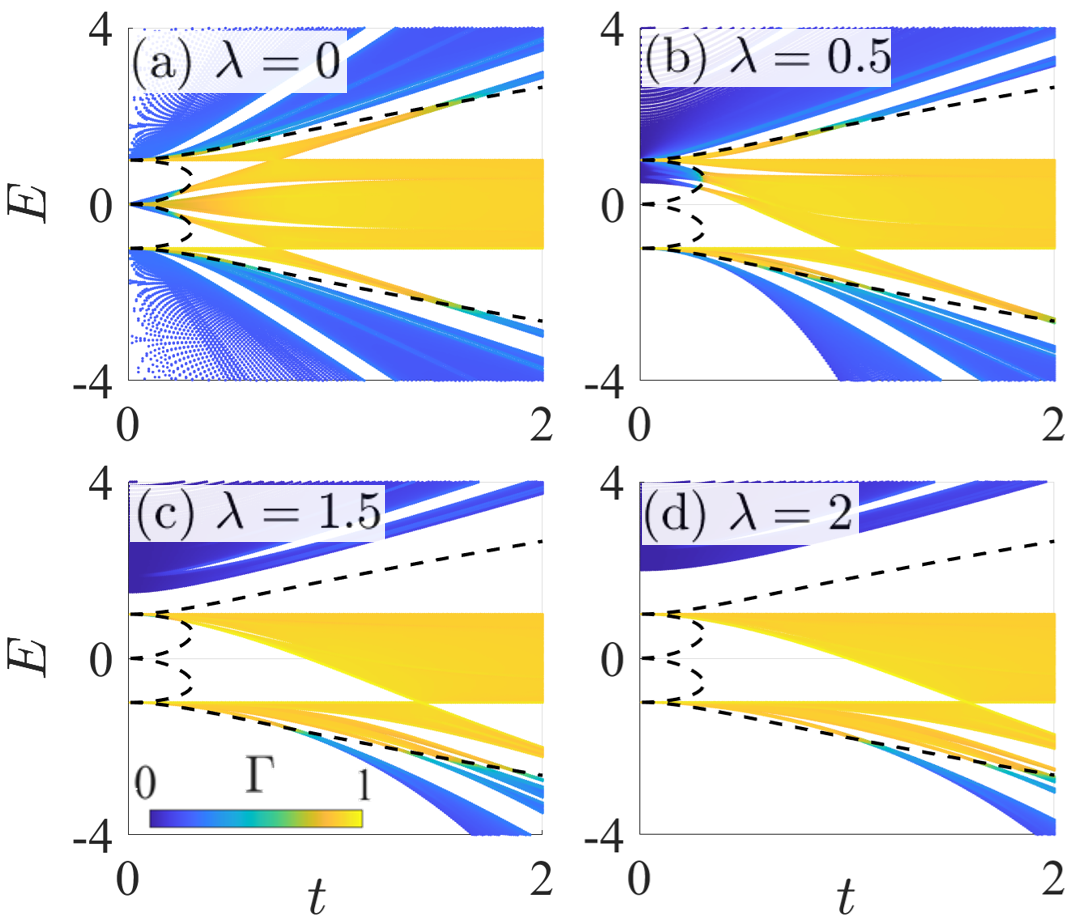}
\caption{Phase diagrams of fractal dimensions in $t$-$E$ plane with $\lambda=0$ (a), $0.5$ (b), $1.5$ (c) and $2$ (d). The other parameters are the same as Fig.~\ref{F1}(a).}
\label{F3}
\end{figure}

Notably, the quasiperiodic potential becomes a constant in the effective mapped model. Consequently, the mobility edge derived in Eq.~\eqref{E11} remains independent of the potential strengthn $\lambda$. Strikingly, the mobility edge persists even when $\lambda=0$ (i.e., in the absence of a quasiperiodic potential). This conclusion agrees with the analytical and numerical results (see Fig.\ref{F3}), which show that while the spatial profiles of localized and extended regions vary with $\lambda$, the mobility edge itself remains invariant.

\section{CONCLUSION}\label{Sec5}

In summary, we have introduced a novel quasiperiodic lattice model characterized by both unbounded hoppings and unbounded potentials. Our results reveal that the unbounded hopping term effectively counteracts the confinement effects induced by the unbounded potential, enabling the formation of extended states. By using Avila's global theory, we derive an exact analytical expression for the mobility edge, which agrees well with the numerical simulations.

Further theoretical analysis demonstrates that the proposed model can be mapped to an effective Aubry-Andr\'{e} model. This mapping unveils an underlying self-duality symmetry in the system, despite the absence of explicit self-duality in the original system. These findings advance the understanding of localization-delocalization transitions in unbounded quasiperiodic systems and provide a useful framework for future studies on unbounded quantum models.

\section{Acknowledgements.}
We thank Yi-Qi Zheng and Zhi-Bing Liang for their insightful suggestions. This work was supported by the National Key Research and Development Program of China (Grant No.2022YFA1405300), and Guangdong Provincial Quantum Science Strategic Initiative(Grant No. GDZX2304002).

J.-M. Zhang and S.-Z. Li contribute equally to this work.


\begin{thebibliography}{99}


\bibitem{PWAnderson1958} P. W. Anderson, Absence of diffusion incertain random lattices,
 \href{https://doi.org/10.1103/PhysRev.109.1492}
{Phys. Rev. \textbf{ 109}, 1492 (1958)}.


\bibitem{EAbrahams1979} E. Abrahams, P. W. Anderson, D. C. Licciardello, and T. V. Ramakrishnan, Scaling Theory of Localization: Absence of Quantum Diffusion in Two Dimensions,
\href{https://doi.org/10.1103/PhysRevLett.42.673}
{Phys. Rev. Lett. \textbf{ 42}, 673 (1979)}.

\bibitem{PALee1985} P. A. Lee and T. V. Ramakrishnan, Disordered electronic systems,
\href{https://doi.org/10.1103/RevModPhys.57.287}
{Rev. Mod. Phys. \textbf{ 57}, 287 (1985)}.

\bibitem{BHetenyi2021} B. Het\'enyi, S. Parlak, and M. Yahyavi, Scaling and renormalization in the modern theory of polarization: Application to disordered systems,
\href{https://doi.org/10.1103/PhysRevB.104.214207}
{Phys. Rev. B \textbf{104}, 214207 (2021)}.


\bibitem{FEvers2008} F. Evers and A. D. Mirlin, Anderson transitions,
\href{https://doi.org/10.1103/RevModPhys.80.1355}
{ Rev. Mod. Phys. \textbf{ 80}, 1355 (2008)}.

\bibitem{ALagendijk2009} A. Lagendijk, B. Tiggelen, and D. S. Wiersma, Fifty years of Anderson localization,
\href{https://doi.org/10.1063/1.3206091}
{Phys. Today \textbf{ 62}, 24 (2009)}.

\bibitem{NMott1987} N. Mott, The mobility edge since 1967,
\href{https://iopscience.iop.org/article/10.1088/0022-3719/20/21/008}
{Journal of Physics C: Solid State Physics \textbf{20}, 3075(1987)}.



\bibitem{RWhitney2014} R. Whitney, Most efficient quantum thermoelectric at finite power output,
\href{https://doi.org/10.1103/PhysRevLett.112.130601}
{ Phys. Rev. Lett. \textbf{112}, 130601 (2014)}.

\bibitem{YamamotoK2017} Yamamoto, A. Aharony, O. Entin Wohlman, and N. Hatano, Thermoelectricity near Anderson localization transitions,
\href{https://doi.org/10.1103/PhysRevB.96.155201}
 {Phys. Rev. B \textbf{96}, 155201 (2017)}.

\bibitem{CChiaracane2020} C. Chiaracane, M. T. Mitchison, A. Purkayastha, G. Haack, and J. Goold, Quasiperiodic quantum heat engines with a mobility edge,
\href{https://doi.org/10.1103/PhysRevResearch.2.013093}
{Phys. Rev. Research, \textbf{2}, 013093 (2020)}.



\bibitem{DJThouless1988} D. J. Thouless, Localization by a Potential with Slowly Varying Period,
\href{https://doi.org/10.1103/PhysRevLett.61.2141}
{Phys. Rev. Lett. \textbf{61}, 2141 (1988)}.

\bibitem{MKohmoto1983} M. Kohmoto, Metal-Insulator Transition and Scaling for Incommensurate Systems,
\href{https://doi.org/10.1103/PhysRevLett.51.1198}
{Phys. Rev. Lett. \textbf{51}, 1198 (1983)}.

\bibitem{MKohmoto2008} M. Kohmoto and D. Tobe, Localization problem in a quasiperiodic system with spin-orbit interaction,
\href{https://doi.org/10.1103/PhysRevB.77.134204}
{Phys. Rev. B \textbf{77}, 134204 (2008)}.

\bibitem{XCai2013} X. Cai, L.-J. Lang, S. Chen, and Y. Wang, Topological Superconductor to Anderson Localization Transition in One-Dimensional Incommensurate Lattices,
\href{https://doi.org/10.1103/PhysRevLett.110.176403}
{Phys. Rev. Lett. \textbf{110}, 176403 (2013)}.

\bibitem{GRoati2008} G. Roati, C. D'Errico, L. Fallani, M. Fattori, C. Fort, M. Zaccanti, G. Modugno, M. Modugno, and M. Inguscio, Anderson localization of a non-interacting Bose-Einstein condensate,
\href{https://doi.org/10.1038/nature07071}
{Nature \textbf{453}, 895 (2008)}.

\bibitem{YLahini2009} Y. Lahini, R. Pugatch, F. Pozzi, M. Sorel, R. Morandotti, N. Davidson, and Y. Silberberg, Observation of a Localization Transition in Quasiperiodic Photonic Lattices,
\href{https://doi.org/10.1103/PhysRevLett.103.013901}
{ Phys. Rev. Lett. \textbf{103}, 013901 (2009)}.

\bibitem{DTanese2014} D. Tanese, E. Gurevich, F. Baboux, T. Jacqmin, A. Lema\^itre, E. Galopin, I. Sagnes, A. Amo, J. Bloch, and E. Akkermans, Fractal energy spectrum of a polariton gas in a Fibonacci quasiperiodic potential,
\href{https://doi.org/10.1103/PhysRevLett.112.146404}
{Phys. Rev. Lett. \textbf{112}, 146404 (2014)}.

\bibitem{HPLuschen2018} H. P. L\"uschen, S. Scherg, T. Kohlert, M. Schreiber, P. Bordia, X. Li, S. D. Sarma, and I. Bloch, Single-particle mobility edge in a one-dimensional quasiperiodic optical lattice,
\href{https://doi.org/10.1103/PhysRevLett.120.160404}
{ Phys. Rev. Lett. \textbf{120}, 160404 (2018)}.

\bibitem{FAAn2018} F. A. An, E. J. Meier, and B. Gadway, Engineering a fluxdependent mobility edge in disordered zigzag chains,
\href{https://doi.org/10.1103/PhysRevX.8.031045}
{ Phys. Rev. X \textbf{8}, 031045 (2018)}.

\bibitem{FAAn2021} F. A. An, K. Padavi\'e, E. J. Meier, S. Hegde, S. Ganeshan, J. H. Pixley, S. Vishveshwara, and B. Gadway, Interactions and mobility edges: Observing the generalized Aubry-Andr\'e model,
\href{https://doi.org/10.1103/PhysRevLett.126.040603}
{Phys. Rev. Lett. \textbf{126}, 040603 (2021)}.

\bibitem{YWang2022a} Y. Wang, J.-H Zhang, Y. Li, J. Wu, W. Liu, F. Mei, Y. Hu, L. Xiao, J. Ma, C. Chin, and S. Jia, Observation of interactioninduced mobility edge in an atomic Aubry-Andr\'e wire,
\href{https://doi.org/10.1103/PhysRevLett.129.103401}
{Phys. Rev. Lett. \textbf{129}, 103401 (2022)}.

\bibitem{HLi2023} H. Li, Y.-Y Wang, Y.-H Shi, K. Huang, X. Song, G.-H Liang, Z.-Y Mei, B. Zhou, H. Zhang, J.-C Zhang, S. Chen, S.-P. Zhao, Y. Tian, Z.-Y Yang, Z. Xiang, K. Xu, D. Zheng, and H. Fan, Observation of critical phase
transition in a generalized Aubry-Andr\'e-Harper model with superconducting circuits,
\href{https://doi.org/10.1038/s41534-023-00712-w}
{npj Quantum Inf. \textbf{9}, 40 (2023)}.

\bibitem{TYLi2024} T. Li, Y. Peng, Y. Wang, H. Hu, Anderson transition and mobility edges on hyperbolic lattices with randomly connected boundaries,
\href{https://doi.org/10.1038/s42005-024-01848-7}
{Commun. Phys. \textbf{7}, 371 (2024)}.




\bibitem{SAubry1980} S. Aubry and G. Andr\'e, Analyticity breaking and Anderson localization in incommensurate lattices, Ann. Israel Phys. Soc \textbf{3}, 18 (1980).


\bibitem{PGHarper1955} P. G. Harper, Single Band Motion of Conduction Electrons in a Uniform Magnetic Field,
\href{https://iopscience.iop.org/article/10.1088/0370-1298/68/10/304}
{Proc. Phys. Soc. A \textbf{68}, 874 (1955)}.

\bibitem{MGon?alves2022} M. Gon?alves, B. Amorim, E. V. Castro, and P. Ribeiro, Hidden dualities in 1d quasiperiodic lattice models,
\href{https://www.scipost.org/10.21468/SciPostPhys.13.3.046?acad_field_slug=physics}
{SciPost Phys. \textbf{13}, 046 (2022)}.




\bibitem{SDSarma1988} S. Das Sarma, S. He, and X.-C. Xie, Mobility edge in a model one-dimensional potential,
\href{https://doi.org/10.1103/PhysRevLett.61.2144}
{Phys. Rev. Lett. \textbf{61}, 2144 (1988)}.

\bibitem{SDSarma1990} S. Das Sarma, S. He, and X.-C. Xie, Localization, mobility edges, and metal-insulator transition in a class of one-dimensional slowly varying deterministic potentials,
\href{https://doi.org/10.1103/PhysRevB.41.5544}
{Phys. Rev. B \textbf{41}, 5544 (1990)}.

\bibitem{ZLu2022} Z. Lu, Z. Xu, and Y. Zhang, Exact Mobility Edges and Topological Anderson Insulating Phase in a Slowly Varying Quasiperiodic Model,
\href{https://doi.org/10.1002/andp.202200203}
{ Ann. Phys. (Berlin) \textbf{534}, 2200203 (2022)}.

\bibitem{YWang2020a} Y. Wang, X. Xu, L. Zhang, H. Yao, S. Chen, J. You, Q. Zhou, and X.-J. Liu, One-Dimensional Quasiperiodic Mosaic Lattice with Exact Mobility Edges,
\href{https://doi.org/10.1103/PhysRevLett.125.196604}
 {Phys. Rev. Lett. \textbf{125}, 196604 (2020)}.

 \bibitem{SLZhu2013} S.-L. Zhu, Z.-D. Wang, Y.-H. Chan, and L.-M. Duan, Topological Bose-Mott Insulators in a One-Dimensional Optical Superlattice,
Phys. Rev. Lett. \textbf{110}, 075303 (2013).

\bibitem{SGaneshan2015} S. Ganeshan, J. H. Pixley, and S. D. Sarma, Nearest Neighbor Tight Binding Models with an Exact Mobility Edge in One Dimension,
\href{https://doi.org/10.1103/PhysRevLett.114.146601}
{Phys. Rev. Lett. \textbf{114}, 146601 (2015)}.

\bibitem{HYao2019} H. Yao, H. Khouldi, L. Bresque, and L. Sanchez-Palencia, Critical behavior and fractality in shallow one-dimensional quasi-periodic potentials,
\href{https://doi.org/10.1103/PhysRevLett.123.070405}
{Phys. Rev. Lett. \textbf{123}, 070405 (2019)}.

\bibitem{XLi2020} X. Li and S. Das Sarma, Mobility edge and intermediate phase in one-dimensional incommensurate lattice potentials,
\href{https://doi.org/10.1103/PhysRevB.101.064203}
{ Phys. Rev. B \textbf{101}, 064203 (2020)}.

\bibitem{TLiu2022} T. Liu, X. Xia, S. Longhi, and L. Sanchez-Palencia, Anomalous mobility edges in one-dimensional quasiperiodic models,
\href{https://doi.org/10.21468/SciPostPhys.12.1.027}
{SciPost Phys. \textbf{12}, 027 (2022)}.

\bibitem{XPLi2016} X.-P. Li, J. H. Pixley, D.-L. Deng, S. Ganeshan, and S. Das Sarma, Quantum nonergodicity and fermion localization in a system with a single-particle mobility edge,
\href{https://doi.org/10.1103/PhysRevB.93.184204}
{Phys. Rev. B \textbf{93}, 184204 (2016)}.

\bibitem{XLi2017} X. Li, X.-P. Li, and S. Das Sarma, Mobility edges in one-dimensional bichromatic incommensurate potentials,
\href{https://doi.org/10.1103/PhysRevB.96.085119}
{Phys. Rev. B \textbf{96}, 085119 (2017)}.

\bibitem{BFZhu2023}B.-F. Zhu, L.-J. Lang, Q Wang, Q.-J. Wang, and Y.-D. Chong, Topological Transitions with an Imaginary Aubry-Andre-Harper Potential,
\href{https://doi.org/10.1103/PhysRevResearch.5.023044}
{Phys. Rev. Res. \textbf{5}, 023044 (2023)}.

\bibitem{EWLiang2023}E.-W. Liang, L.-Z. Tang, and D.-W. Zhang, Quantum criticality and Kibble-Zurek scaling in the Aubry-Andr\'e-Stark model,
\href{https://arxiv.org/abs/2405.10199}
{arXiv:2405.10199}.




\bibitem{JBiddle2010} J. Biddle and S. Das Sarma, Predicted Mobility Edges in One-Dimensional Incommensurate Optical Lattices: An Exactly Solvable Model of Anderson Localization,
\href{https://doi.org/10.1103/PhysRevLett.104.070601}
{Phys. Rev. Lett. \textbf{104}, 070601 (2010)}.

\bibitem{JBiddle2011} J. Biddle, D. J. Priour, B. Wang, and S. Das Sarma, Localization in one-dimensional lattices with non-nearest-neighbor hopping: Generalized Anderson and Aubry-Andr\'e models,
\href{https://doi.org/10.1103/PhysRevB.83.075105}
{Phys. Rev. B \textbf{83}, 075105 (2011)}.

\bibitem{XDeng2019} X. Deng, S. Ray, S. Sinha, G. V. Shlyapnikov, and L. Santos, One-Dimensional Quasicrystals with Power-Law Hopping,
\href{https://doi.org/10.1103/PhysRevLett.123.025301}
{Phys. Rev. Lett. \textbf{123}, 025301 (2019)}.

\bibitem{XXia2022} X. Xia, K. Huang, S. Wang, and X. Li, Exact mobility edges in the non-Hermitian t1-t2 model: Theory and possible experimental realizations,
\href{https://doi.org/10.1103/PhysRevB.105.014207}
{Phys. Rev. B \textbf{105}, 014207 (2022)}.

\bibitem{MGon2023} M. Gon?calves, B. Amorim, E. Castro, and P. Ribeiro, Critical phase dualities in 1D exactly solvable quasiperiodic models,
\href{https://doi.org/10.1103/PhysRevLett.131.186303}
{Phys. Rev. Lett. \textbf{131}, 186303 (2023)}.

\bibitem{XCZhou2023a} X.-C. Zhou, Y. Wang, T.F.J. Poon, Q. Zhou, and X.-J. Liu, Exact new mobility edges between critical and localized states,
\href{https://doi.org/10.1103/PhysRevLett.131.176401}
{Phys. Rev. Lett. \textbf{131}, 176401 (2023)}.




\bibitem{YJZhao2025}Y.-J. Zhao, H.-Z. Li, X.-Y. Huang, S.-Z. Li, and J.-X. Zhong, Fate of pseudomobility edges and multiple states in a non-Hermitian Wannier-Stark lattice,
\href{https://doi.org/10.1103/PhysRevB.111.014315}
{Phys. Rev. B, \textbf{111}, 014315(2025)}.

\bibitem{SZLi2024a}S.-Z. Li, E.-H. Cheng, S.-L. Zhu, and Z. Li, Asymmetric transfer matrix analysis of Lyapunov exponents in one-dimensional nonreciprocal quasicrystals,
\href{https://doi.org/10.1103/PhysRevB.110.134203}
{Phys. Rev. B \textbf{110}, 134203(2024)}.

\bibitem{SZLi2024b}S.-Z. Li and Z. Li, Ring structure in the complex plane: A fingerprint of a non-Hermitian mobility edge,
\href{https://doi.org/10.1103/PhysRevB.110.L041102}
{Phys. Rev. B \textbf{110}, L041102(2024)}.

\bibitem{GJLiu2024}G.-J. Liu, J.-M. Zhang, S.-Z. Li, and Z. Li, Emergent strength-dependent scale-free mobility edge in a nonreciprocal long-range Aubry-Andr\'e-Harper model,
\href{https://doi.org/10.1103/PhysRevA.110.012222}
{Phys. Rev. A \textbf{110}, 012222(2024)}.

\bibitem{SZLi2024c}S.-Z. Li, X.-J. Yu, and Z. Li, Emergent entanglement phase transitions in non-Hermitian Aubry-Andr\'e-Harper chains,
\href{https://doi.org/10.1103/PhysRevB.109.024306}
{Phys. Rev. B \textbf{109}, 024306(2024)}.

\bibitem{SLJiang2023}S.-L. Jiang, Y.-X. Liu, and L.-J. Lang, General mapping of one-dimensional non-Hermitian mosaic models to non-mosaic counterparts: Mobility edges and Lyapunov exponents,
\href{https://doi.org/10.1088/1674-1056/ace426}
{Chin. Phys. B \textbf{32}, 097204 (2023)}.

\bibitem{DWZhang2020a}D.-W. Zhang, Y.-L. Chen, G.-Q. Zhang, L.-J. Lang, Z. Li, and S.-L. Zhu, Skin superfluid, topological Mott insulators, and asymmetric dynamics in interacting non-Hermitian Aubry-Andre-Harper models,
\href{https://doi.org/10.1103/PhysRevB.101.235150}
{Phys. Rev. B \textbf{101}, 235150 (2020)}.

\bibitem{DWZhang2020b}D.-W. Zhang, L.-Z. Tang, L.-J. Lang, H. Yan, and S.-L. Zhu, Non-Hermitian Topological Anderson Insulators,
\href{https://doi.org/10.1007/s11433-020-1521-9}
{Sci. China-Phys. Mech. Astron. \textbf{63}, 267062 (2020)}.

\bibitem{HJiang2019}H. Jiang, L.-J. Lang, C. Yang, S.-L. Zhu, and S. Chen, Interplay of non-Hermitian skin effects and Anderson localization in non-reciprocal quasiperiodic lattices,
\href{https://doi.org/10.1103/PhysRevB.100.054301}
{Phys. Rev. B 100, 054301 (2019)}.

\bibitem{JLDong2025}J.-L. Dong, E.-W. Liang, S.-Y. Liu, G.-Q. Zhang, L.-Z. Tang, and D.-W. Zhang, Interplay of non-Hermitian skin effects and Anderson localization in non-reciprocal quasiperiodic lattices,
\href{https://arxiv.org/abs/2501.03777}
{arXiv:2501.03777}.

\bibitem{QLin2022} Q. Lin, T. Li, L. Xiao, K. Wang, W. Yi and P. Xue, Observation of non-Hermitian topological Anderson insulator in quantum dynamics,
\href{https://doi.org/10.1038/s41467-022-30938-9}
{Nat Commun \textbf{13}, 3229 (2022)}.


\bibitem{TLi2022} T. Li, Y.-S. Zhang and W. Yi, Engineering Dissipative Quasicrystals,
\href{https://doi.org/10.1103/PhysRevB.105.125111}
{Phys. Rev. B \textbf{105}, 125111 (2022).}






\bibitem{YHatsugai1990} Y. Hatsugai and M. Kohmoto, Energy spectrum and the quantum Hall effect on the square lattice with nextnearest-neighbor hopping,
\href{https://doi.org/10.1103/PhysRevB.42.8282}
{ Phys. Rev. B \textbf{42}, 8282 (1990)}.

\bibitem{JHHan1994} J.-H. Han, D. J. Thouless, H. Hiramoto, and M. Kohmoto, Critical and bicritical properties of Harper's equation with next-nearest-neighbor coupling,
\href{https://doi.org/10.1103/PhysRevB.50.11365}
{Phys. Rev. B \textbf{50}, 11365(1994)}.

\bibitem{YTakada2004} Y. Takada, K. Ino, and M. Yamanaka, Statistics of spectra for critical quantum chaos in one-dimensional quasiperiodic systems,
\href{https://doi.org/10.1103/PhysRevE.70.066203}
{Phys. Rev. E \textbf{70}, 066203 (2004)}.

\bibitem{FLiu2015} F. Liu, S. Ghosh, and Y.-D. Chong, Localization and adiabatic pumping in a generalized Aubry-Andr\'e-Harper model,
\href{https://doi.org/10.1103/PhysRevB.91.014108}
{ Phys. Rev. B \textbf{91}, 014108 (2015)}.

\bibitem{JWang2016} J. Wang, X.-J. Liu, X. Gao, and H. Hu, Phase diagram of a non-Abelian Aubry-Andr\'e-Harper model with p-wave superfluidity,
\href{https://doi.org/10.1103/PhysRevB.93.104504}
{Phys. Rev. B \textbf{93}, 104504 (2016)}.



\bibitem{SZLi2025} S.-Z. Li, Y.-C. Zhang, Y.-C. Wang, S.C. Zhang, S.-L. Zhu, and Z. Li, Multifractal-enriched mobility edges and emergent quantum phases in one-dimensional exactly solvable lattice models,
\href{https://arxiv.org/abs/2501.07866}
{arXiv:2501.07866}.



\bibitem{SJitomirskaya2017} S. Jitomirskaya and F. Yang, Singular Continuous Spectrum for Singular Potentials, \href{https://link.springer.com/article/10.1007/s00220-016-2823-4}
{Commun. Math. Phys. \textbf{351}, 1127 (2017)}.

\bibitem{FYang2019} F. Yang and S. Zhang, Singular continuous spectrum and generic full spectral/packing dimension for unbounded quasiperiodic Schr?dinger operators,
\href{https://doi.org/10.1007/s00023-019-00810-6}
{ Ann. Henri Poincar \textbf{20}, 2481 (2019)}.

\bibitem{JWang2022} T. Liu, X. Xia, S. Longhi, and L. Sanchez-Palencia, Anomalous mobility edges in one-dimensional quasiperiodic models,
\href{https://www.scipost.org/10.21468/SciPostPhys.12.1.027?acad_field_slug=physics}
{ Sci Post Phys. \textbf{12}, 027 (2022)}.

\bibitem{YCZhang2022} Y.-C. Zhang and Y.-Y. Zhang, Lyapunov exponent, mobility edges, and critical region in the generalized Aubry-Andre model with an unbounded quasiperiodic potential,
\href{https://doi.org/10.1103/PhysRevB.105.174206}
{Phys.Rev.B \textbf{105}, 174206 (2022)}.

\bibitem{XPJiang}X.-P. Jiang, W.-l. Zeng, Y.-Y. Hu, and L. Pan, Exact anomalous mobility edges in one-dimensional non-Hermitian quasicrystals,
\href{https://arxiv.org/abs/2409.03591}
{arXiv:2409.03591}



\bibitem{BSimon1989} B. Simon and T. Spencer, Trace class perturbations and the absence of absolutely continuous spectra, Commun.
\href{https://projecteuclid.org/journals/communications-in-mathematical-physics/volume-125/issue-1/Trace-class-perturbations-and-the-absence-of-absolutely/cmp/1104179386.pdf}
{ Math. Phys. \textbf{125}, 113 (1989)}.

\bibitem{SLonghi2023}S. Longhi, Absence of mobility edges in mosaic Wannier-Stark lattices,
\href{https://doi.org/10.1103/PhysRevB.108.064206}
{Phys. Rev. B \textbf{108},064206 (2023)}.


















\bibitem{YWang2022b}Y. Wang, L. Zhang, W. Sun, T.F.J. Poon, and X.-J. Liu, Quantum phase with coexisting localized, extended, and critical zones,
\href{https://doi.org/10.1103/PhysRevB.106.L140203}
{Phys. Rev. B \textbf{106}, L140203 (2022)}.




\bibitem{APadhan2022} A. Padhan, M. Giri, S. Mondal, and T. Mishra, Emergence of multiple localization transitions in a one-dimensional quasiperiodic lattice,
\href{https://doi.org/10.1103/PhysRevB.105.L220201}
{Phys. Rev. B 105, L220201 (2022)}.

\bibitem{YZhang2022} Y. Zhang, B. Zhou, H. Hu, and S. Chen, Localization, multifractality, and many-body localization in periodically kicked quasiperiodic lattices,
\href{https://doi.org/10.1103/PhysRevB.106.054312}
{Phys. Rev. B 106, 054312 (2022)}.

\bibitem{MSarkar2021} Madhumita Sarkar, R. Ghosh, A. Sen, and K. Sengupta, Mobility edge and multifractality in a periodically driven Aubry-Andr\'e model,
\href{https://doi.org/10.1103/PhysRevB.103.184309}
{Phys. Rev. B 103, 184309 (2021)}.


\bibitem{RQi2023} R. Qi, J.-P. Cao, and X.-P. Jiang, Multiple localization transitions and novel quantum phases induced by a staggered on-site potential,
\href{https://doi.org/10.1103/PhysRevB.107.224201}
{ Phys. Rev. B, 107, 224201 (2023)}.



\bibitem{Avila2015} A. Avila, Global theory of one-frequency Schrodinger operators,
\href{https://projecteuclid.org/journals/acta-mathematica/volume-215/issue-1/Global-theory-of-one-frequency-Schr\%C3\%B6dinger-operators/10.1007/s11511-015-0128-7.full}
{Acta Math. \textbf{215}, 1 (2015)}.

\bibitem{MGong2024}H.-T. Hu, X.-S. Lin, A.-M. Guo, Z.-J. Lin, and M. Gong, Hidden self-duality in quasiperiodic network models,
\href{https://arxiv.org/abs/2411.06843}
{arXiv:2411.06843}.





\end{thebibliography}
\end{document}